\documentclass[12pt,thmsa]{article}
\usepackage{amsfonts}
\usepackage{graphicx}
\usepackage{epsfig}
\usepackage{graphics}

\makeatletter
\newcommand{\row}[1]%
{\mathord{\buildrel{\lower3pt%
\hbox{$\scriptscriptstyle\rightarrow$}}\over #1}}
\newcommand{\col}[1]{{#1^{\raisebox{2pt}[\height]%
{$\scriptstyle\downarrow$}}}}
\newcommand{\dyadic}[1]{\mathord{\dyadic@rrow{#1}}}
\newcommand{\dyadic@rrow}[1]{
\begin{picture}(12,12)(-1,0)
\put(-2,12){\makebox(0,0)[t]{$\scriptscriptstyle\downarrow$}}
\put(-2,12){\makebox(0,0)[l]{$\scriptscriptstyle\longrightarrow$}}
\put(5,0){\makebox(0,0)[b]{$#1$}}
\end{picture}
}

\newcommand{\bra}[1]{\bigl\langle #1 \bigr|}
\newcommand{\ket}[1]{\bigl| #1 \bigr\rangle}

\newcommand{\plA}[3]{Phys.\ Lett.\ \textbf{A#1}, #2 (#3)}

\begin{document}

\begin{center}
{\Large  Information loss  in local dissipation environments}

~

\small{Nasser Metwally}

{\footnotesize Mathematics Department, College of Science, Bahrain
University, 32038 Bahrain }
\end{center}

\begin{abstract}
 The sensitivity of entanglement to the thermal and squeezed  reservoirs' parameters
 is investigated regarding  entanglement decay and what is called
sudden-death of entanglement, ESD, for a system of two qubit
pairs.
 The dynamics of information is investigated by means of
the information disturbance and exchange information. We show that
for squeezed reservoir, we can keep both of the entanglement and
information survival for a long time. The sudden death of
information is seen in the case of  thermal reservoir.

\end{abstract}
\section{Introduction}
Entangled qubits are one of the most promising candidates for
quantum communication and computation. There are many interesting
applications based on these entangled systems. Among these
applications, dense coding \cite{ben1},quantum teleportation
\cite{ben2}, quantum cryptography \cite{ben3} etc. These entangled
systems can not be isolated from their surrounding environments.
 cause deterioration of entanglement. So, investigating and quantifying the amount of
entanglement contained in entangled system interacting with open
systems  is very important in the context of quantum information
\cite{khasin}.  As an example, Yu and Eberly \cite{Eberly1,
Eberly2}, has investigated the dynamics of entanglement for
entangled qubit pairs undergoing various modes of decoherence.
They showed that the dynamics of entanglement between a two  qubit
system interacting independently with classical or quantum noise,
displays two different types of behavior; the phenomena of
entanglement decay and entanglement sudden death, ESD
\cite{Eberly2, Eberly3}. Since  then,  different systems have been
investigated. For some systems, the ESD appears whenever the
system is open or closed \cite{cui}. The effect of the local
squeezed reservoir on initially entangled two qubit system  is
investigated, where it is shown that the squeezing causes
different behavior of entanglement decay on different time scales
\cite{Zhou}. Recently  in \cite{Zubairy}, it has been shown that
the ESD always exists with  thermal  and squeezed reservoirs,
where the authors presented explicit expression for the ESD for
some entangled states. Also, it has been recently shown
\cite{metwally3}, that the ESD and entanglement decay phenomena
appear for qubit system passed through Bloch channel
\cite{Ban1,Ban2}. The ESD under the effect of the individual
environments has been  experimentally seen \cite{Lastra,Almeida}.

 The purpose of this paper is to continue investigating
questions of this sort: by how much are  the entanglement and the
information of entangled  two  qubits  state degrade when it
passes through a thermal or squeezed reservoir.

The paper is organized as follows: In sec.2, we present the model
and its solution. The dynamics of entanglement is investigated in
Sec.3, where we consider two  classes of entangled states as an
initial states, maximum and partial entangled states. Sec.4 is
devoted to study the dynamics of quantum information, where we
quantify the amount of disturbance and exchange information
between the entangled state and the local reservoirs. Finally we
give a conclusion of our results in Sec.5.

\section{The model and its solution}
Assume that we have  a source to  generate entangled qubit pairs.
One qubit is sent to Alice and the other to Bob. The general two
qubit state is defined by
\begin{equation}\label{int}
\rho_{ab}(0)=\frac{1}{4}(1+\row{s}\cdot\col{\sigma_1}+
\row{t}\cdot \col{\sigma_2}+\row{\sigma_1}\cdot
\dyadic{C}\cdot\col{\sigma_2}),
\end{equation}
where the vectors $\row{s}$ and $\row{t}$ are the Bloch vectors
for Alice and Bob's qubits  respectively, $\dyadic{C}$   is a $3
\times 3 $ matrix represents the cross dyadic and $\row{\sigma_1}$
and $\row{\sigma_2}$  are the spin Pauli  vectors \cite
{metwally}. Now assume that each qubit interacts individually with
its  squeezed vacuum environment\cite{Zhou}. Within Markov
approximation, we can write the master equation in the
Schr\"{o}dinger form as,
\begin{equation}\label{Sch}
\frac{\partial}{\partial
t}\rho_{ab}=L_a(\rho_{ab})+L_b(\rho_{ab}),
\end{equation}
with,
\begin{eqnarray}
L_{i}(\rho_{ab})&=&-\frac{\Gamma_i}{2}(1+\mathcal{N}_i)(\sigma_i^+\sigma_i^-\rho_{ab}
-2\sigma_i^-\rho_{12}\sigma_i^++\rho_{ab}\sigma_i^+\sigma_i^-)
\nonumber\\
&&-\frac{\Gamma_i}{2}\mathcal{N}_i(\sigma_i^-\sigma_i^+\rho_{ab}
-2\sigma_i^+\rho_{12}\sigma_i^-+\rho_{ab}\sigma_i^-\sigma_i^+)
\nonumber\\
&&-\frac{\Gamma_i}{2}\mathcal{M}_i(\sigma_i^+\sigma_i^+\rho_{ab}
-2\sigma_i^+\rho_{12}\sigma_i^++\rho_{ab}\sigma_i^+\sigma_i^+)
\nonumber\\
&&-\frac{\Gamma_i}{2}\mathcal{M}^*_i(\sigma_i^-\sigma_i^-\rho_{ab}
-2\sigma_i^-\rho_{12}\sigma_i^-+\rho_{ab}\sigma_i^-\sigma_i^-),
\end{eqnarray}
where $i=1,2$ refers to the first (Alice's qubit) and the second
for (Bob's qubit), $\Gamma_i$ is the atomic spontaneous emission
rate in local squeezed field. The parameter
$\mathcal{M}=|\mathcal{M}|e^{i\theta}$, describes the strength of
the two photon correlation, where
$|\mathcal{M}_i|\leq\sqrt{\mathcal{N}_i(1+\mathcal{N}_i)}$.
Finally, $\sigma^\pm_i=\sigma_{ix}\pm i\sigma_{iy}$.

To investigate the dynamical behavior of the initial entangled
state $\rho_{ab}(0)$, we solve the Schr\"{o}dinger equation
(\ref{Sch}). In this context, we use the Kraus representation
described in \cite{Zhou}. The time-evolution of the input state
(\ref{int}) is given by
\begin{equation}\label{tot}
\rho_{ab}(t)=\sum_j^{4}
{\kappa_j^a\otimes\kappa_j^b\rho_{ab}(0)\kappa_j^{a\dagger}\kappa_j^{b\dagger}}.
\end{equation}
 For our analysis, we describe the Kraus operators in the
computational basis $\ket{0}$ and $\ket{1}$ as,
\begin{eqnarray}\label{krus}
\kappa_1^{i}&=&\alpha_1^{i}\ket{0}\bra{1}+\beta_1^{i}\ket{1}\bra{1},\quad
\kappa_2^{i}=\beta_2^{i}\ket{1}\bra{1},
\nonumber\\
\kappa_3^i&=&\alpha_3^i\ket{0}\bra{1}+\beta_3^{i}\ket{1}\bra{0},\quad
\kappa_4^i=\alpha_4^i\ket{1}\bra{0},
\end{eqnarray}
 where
 \begin{eqnarray}\label{parm}
 \alpha_1^i&=&e^{-\frac{\zeta t}{2}}\sqrt{\cosh\zeta t+\frac{\Gamma_i}{2\zeta}\sinh\zeta
 t}, \quad \beta_1^i=e^{-\frac{\zeta t}{2}}\frac{\cosh\eta
 t}{\alpha_1^i},
 \nonumber\\
 \beta_2^i&=&e^{-\frac{\zeta t}{2}}\sqrt{\frac{[1-(\frac{\Gamma_i}{2\zeta})^2]
 \sinh^2\zeta t-\sinh^2\eta t}
 {\cosh \zeta t +\frac{\Gamma_i}{2\zeta}\sinh \zeta t}},
\nonumber\\
\alpha_3^i&=&e^{-\frac{\zeta t}{2}}\frac{\sinh(\eta
t)}{\sqrt{(1+\frac{\Gamma_i}{2\zeta})\sinh(\eta t)}},\quad
\beta_3^i=e^{-\frac{\zeta
t}{2}}\sqrt{(1+\frac{\Gamma_i}{2\eta})\sinh(\eta t)}~e^{-i\theta},
\nonumber\\
\alpha_4^i&=&e^{-\frac{\zeta t}{2}}
\sqrt{\frac{[1-(\frac{\Gamma_i}{2\zeta})^2]\sinh^2(\zeta t)-\sinh
^2(\eta t)}
 {(1+\frac{\Gamma_i}{2\eta})\sinh(\eta t)}},
 \end{eqnarray}
 $\zeta=\frac{\Gamma_i}{2}(2\mathcal{N}_i+1)$ and
$\eta=\Gamma_i|\mathcal{M}_i|$.

To show our idea, let us  consider   a class of entangled states
with zero Bloch vectors, $(\row{s}=\row{t}=0)$. This
simplification leads us to what is called a generalized Werner
state \cite{Wer,Metwally2},
\begin{equation}\label{ini}
\rho(0)=\frac{1}{4}(1+c_1\sigma_x^{(1)}\sigma_x^{(2)}+c_2\sigma_y^{(1)}\sigma_y^{(2)}
+c_3\sigma_z^{(1)}\sigma_z^{(2)}).
\end{equation}
By  means of Bell states
$\ket{\phi^\pm}=\frac{1}{\sqrt{2}}(\ket{00}\pm\ket{11})$ and
$\ket{\psi^\pm}=\frac{1}{\sqrt{2}}(\ket{01}\pm \ket{10})$, we  can
rewrite this initial state (\ref{ini}) as,
\begin{eqnarray}\label{ini1}
\rho(0)&=&\frac{1+c_1+c_2+c_3}{4}\ket{\phi^+}\bra{\phi^+}
+\frac{1-c_1+c_2+c_3}{4}\ket{\phi^-}\bra{\phi^-}
 \nonumber\\
&+&\frac{1-c_1-c_2-c_3}{4}\ket{\psi^-}\bra{\psi^-}+
\frac{1+c_1+c_2-c_3}{4}\ket{\psi^+}\bra{\psi^+}.
\end{eqnarray}
From this class of states we can get the singlet state
$(\ket{\psi^-}\bra{\psi^-})$ if we set $c_1=c_2=c_3=-1$, and if
$c_1=c_2=c_3=1$, one gets $\ket{\phi^+}\bra{\phi^+}$ and so on.
Also, if we assume that $c_1=c_2=c_3=x$, one gets  Werner state
\cite{Wer},
\begin{equation}
\rho_w(0)=\frac{3x+1}{4}\ket{\psi^-}\bra{\psi^-}+
\frac{1-x}{4}\left(\ket{\phi^+}\bra{\phi^+}+\ket{\phi^-}\bra{\phi^-}+
\ket{\psi^+}\bra{\psi^+}\right).
\end{equation}
Now, let us assume that we have a source  which supplies us with
an entangled  qubit state of form (\ref{ini1}). The qubits leave
each other and then interact with their local reservoir. With the
Kraus operators, the time evolvement of the  density operator
(\ref{ini1}) is,
\begin{eqnarray}\label{out}
\rho(t)&=&e^{-i\Gamma
t}\Big[\frac{1+c_3}{8}\Bigl\{(s_1+s_4)(\ket{\phi^+}\bra{\phi^+}+\ket{\phi^-}\bra{\phi^-})
+(s_1-s_4)(\ket{\phi^+}\bra{\phi^-}+\ket{\phi^-}\bra{\phi^+})\Bigr\}
\nonumber\\
&+&\frac{c_1-c_2}{8}\Bigl\{(s_2+s_3)(\ket{\phi^+}\bra{\phi^+}-\ket{\phi^-}\bra{\phi^-})
+(s_2-s_3)(\ket{\phi^-}\bra{\phi^+}-\ket{\phi^+}\bra{\phi^-})\Bigr\}
\nonumber\\
&+&\frac{1-c_3}{8}\Bigl\{(s_5+s_8)(\ket{\psi^+}\bra{\psi^+}+\ket{\psi^-}\bra{\psi^-})
+(s_5-s_8)(\ket{\psi^+}\bra{\psi^-}+\ket{\psi^-}\bra{\psi^+})\Bigr\}
\nonumber\\
&+&\frac{c_1+c_2}{8}\Bigl\{(s_6+s_7)(\ket{\psi^+}\bra{\psi^+}-\ket{\psi^-}\bra{\psi^-})
+(s_6-s_7)(\ket{\psi^-}\bra{\psi^+}-\ket{\psi^+}\bra{\psi^-})\Bigr\}\Bigr],
\nonumber\\
\end{eqnarray}
where,
\begin{eqnarray}
s_1&=&\sum_{i=1,3}{|\alpha_i^a|^2|\alpha_i^b|^2},\quad
s_2=\sum_{i=1,3}{\alpha_i^a\alpha_i^b\beta_i^{*a}\beta_i^{*b}},\quad\quad
s_3=\sum_{i=1,3}{\beta_i^a\beta_i^b\alpha_i^{*a}\alpha_i^{*b}},
\nonumber\\
s_4&=&\sum_{i=2,4}{|\alpha_i^a|^2|\alpha_i^b|^2}+\sum_{i=1,3}{|\beta_i^a|^2|\beta_i^b|^2},
\quad \hspace{1.8cm}
s_5=\sum_{i=1,3}{|\alpha_i^a|^2|\beta_i^b|^2},
\nonumber\\
s_6&=&\sum_{i=1,3}{\alpha_i^a\beta_i^b\beta_i^{*a}\alpha_i^{*b}},\quad
s_7=\sum_{i=1,3}{\beta_i^a\alpha_i^b\alpha_i^{*a}\beta_i^{*b}},
\quad s_8=\sum_{i=1,3}{|\beta_i^a|^2|\alpha_i^b|^2},
\nonumber\\
\Gamma&=&\Gamma_1+\Gamma_2.
\end{eqnarray}

\section{ Entanglement Dynamics }
In this section, we investigate  the robustness of the entangled
state  when each qubit interacts with its own reservoir
individually, where we study the effect   of reservoir parameters.
 In our {\it first example}, we assume that the source supplies
the user Alice and Bob with maximum entangled state say,
$\ket{\phi^+}\bra{\phi^+}$ with $c_1=c_2=c_3=1$ or partially
entangled state with  $c_1=c_2=c_3=0.85$. Unfortunately, each
qubit forced to passes through individual reservoir for some time.
During this time there is non desirable interactions between the
qubits and the reservoir. These interactions cause deteriorate of
the amount entanglement contained in the entangled two qubits
state and consequently the efficiency of performing quantum
information tasks decreases. In our treatment we consider the
local reservoirs  to be thermal or squeezed.
\begin{figure}[b!]
  \begin{center}
\includegraphics[width=18pc,height=12pc]{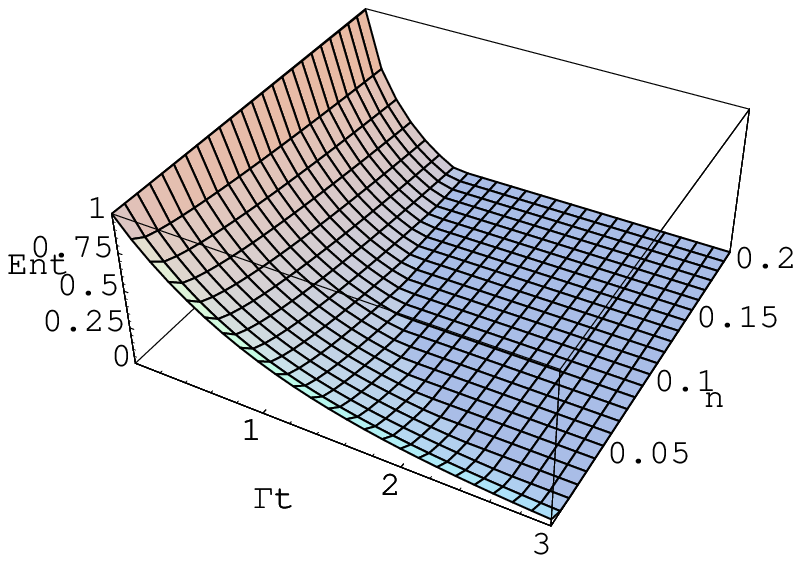}
\includegraphics[width=18pc,height=12pc]{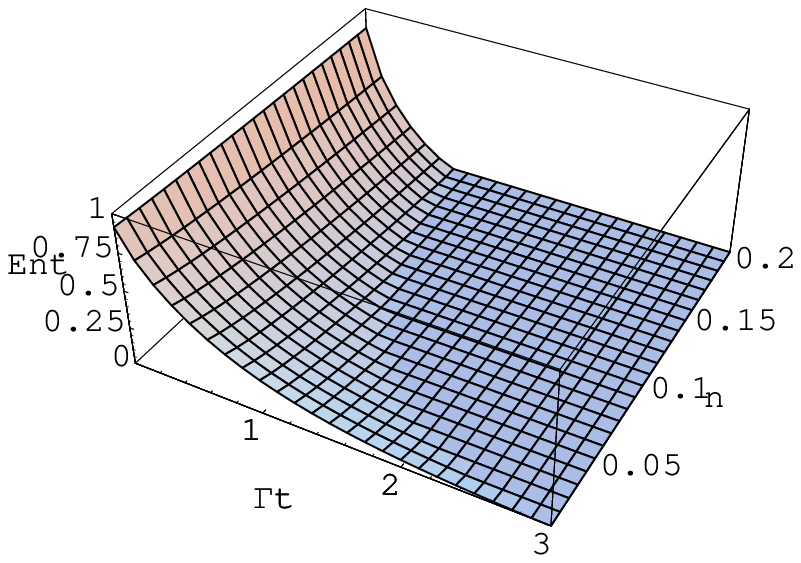}\
\put(-400,10){(a)} \put(-100,8){(b)}\
    \caption{The effect of the mean photon number $\mathcal{N}_1=\mathcal{N}_2=n$
   on the degree of entanglement(a)
   The initial state is maximum entangled state
    $\ket{\phi^+}\bra{\phi^+}$ and (b) For a partial entangled state
     with $c_1=c_2=c_3=0.85$. }
  \end{center}
\end{figure}

To  quantify the degree of entanglement we use the negativity,
 where it is  easily calculable measure.
 It  is given in terms of the eigenvalues of the partial transpose of the
 density operator  \cite{Kyc},
 \begin{equation}
 DoE=\sum_{i}{|\lambda_i|-1},
 \end{equation}
where $\lambda_i$ are the eigenvalues of the partial transpose of
the output density operator (\ref{out}).

Fig.$1$, shows the dynamics of entanglement against  the
normalized time $\Gamma t$, where we assume that the two qubits
pass through a thermal reservoir, i.e
$\mathcal{M}_1=\mathcal{M}_2=0$ and we assume that the two
reservoirs  have the same number of photons,
$\mathcal{N}_1=\mathcal{N}_2= n$. In Fig.(1a), we consider the
case where the source supplies the partners Alice and Bob with
maximum entangled state.
 It is clear that for small values of $n$,
we can see that the entanglement decays asymptotically  and the
time of the sudden death is  delayed.
  For large values of $n$
 the decay of entanglement is hastened and the time of entanglement sudden death
 becomes shorter.

In Fig.(1b), we quantify the amount of survival amount of
entanglement contained in  a density operator initially prepared
in a partially entangled state. It is clear that, the entanglement
decays faster and  the time of the entanglement sudden death is
shorter. From Figs.(1a) and Fig.(1b), we  see that the
entanglement decay and the sudden death of entanglement  are
sensitive to the initial entangled state.
 So, by controlling  $n$ and $\Gamma$ one  can prolonge the
time of lived entanglement and delayed the time of the sudden
death .
\begin{figure}[b!]
  \begin{center}
  \includegraphics[width=18pc,height=12pc]{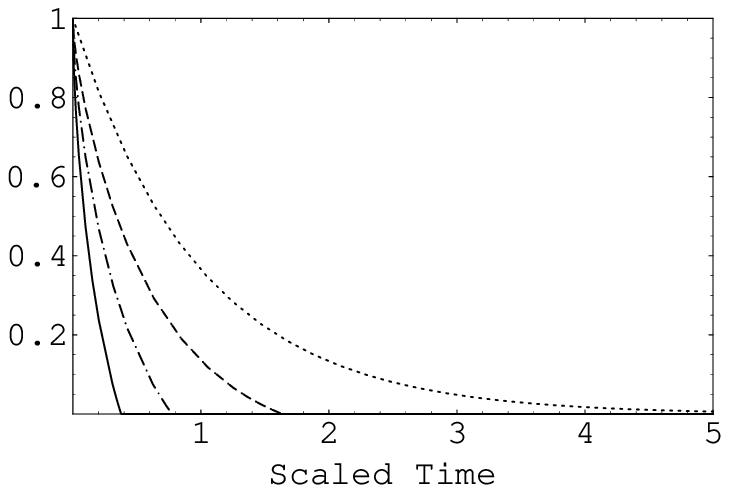}\
  \includegraphics[width=18pc,height=12pc]{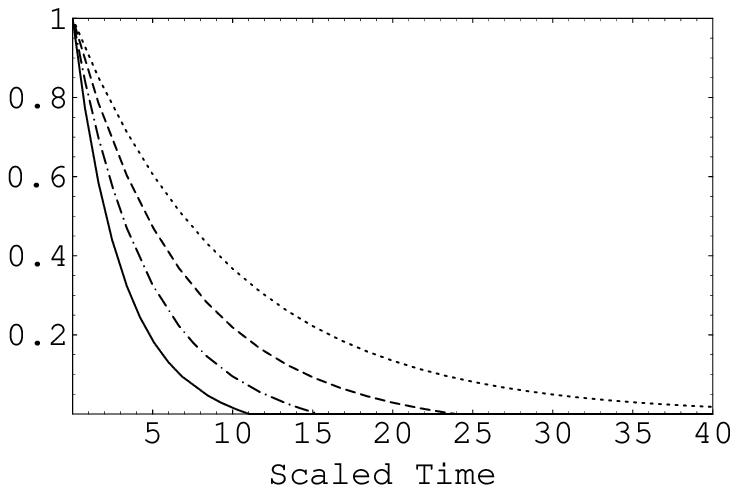}
  \put(-220,80){DoE}
 \put(-430,80){DoE}
      \caption{The  dot, dash-dot, dash and slid curves represent
      the degree of entanglement(a) inside a thermal reservoir
     $\mathcal{N}_1=\mathcal{N}_2=n=0.00001, 0.05, 0.2, 0.6$ and
     (b)inside a squeezed reservoir with
     $\mathcal{M}_1=\mathcal{M}_2=0.05,0.2,0.4,0.6$ and }
  \end{center}
\end{figure}

In Fig.(2a), we investigate the dynamics of entanglement  at some
specific values of the mean photon number, $n$, where we assume
that the qubits interact with thermal reservoir. For small values
of $n=0.00001$, the phenomena of long-lived entanglement is seen,
where the entanglement decreases asymptotically. On the other
hand, as $n$  increases,  the time at which the entanglement
vanishes decreases. The entanglement sudden death phenomena
appears for much larger values of the mean photon number, $n=6$.
The behavior of the entanglement when the qubits interact with
squeezed reservoir  is depicted in Fig.(2a).  In this case the
entanglement decays smoothly and for small values of the squeezed
parameter, $\mathcal{M}$, one observes the long lived
entanglement. Also, the vanishing time of entanglement is much
larger than that has been shown in Fig.(2a).

\section{ Information  dynamics }
Quantum information science has emerged as one of the most
exciting scientific developments in the past decade.
 Let us assume that Alice and Bob have coded information, say a quantum secret
key, in their shared entangled state. But due to the reality there
is no isolated systems, the entangled state which carries the
information interacts with its surroundings. These undesirable
interactions cause a  loss of information. Therefore quantum
information, in fact, cannot be perfectly copied, neither locally
\cite{Gisn} nor at distance\cite{van}.

Our aim in this section is to  investigate the dynamics of
information which is carried by the shared entangled state. In our
calculations, we consider the effect of the thermal and the
squeezed reservoirs. In this treatments, we investigate two
phenomena, the disturbance of information and the exchange
information between the shared state and its local environments
\begin{figure}[b!]
  \begin{center}
  \includegraphics[width=18pc,height=12pc]{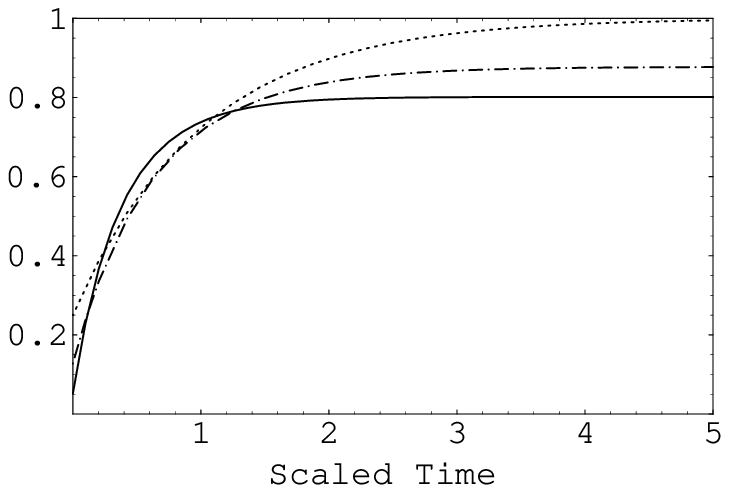}
 \includegraphics[width=18pc,height=12pc]{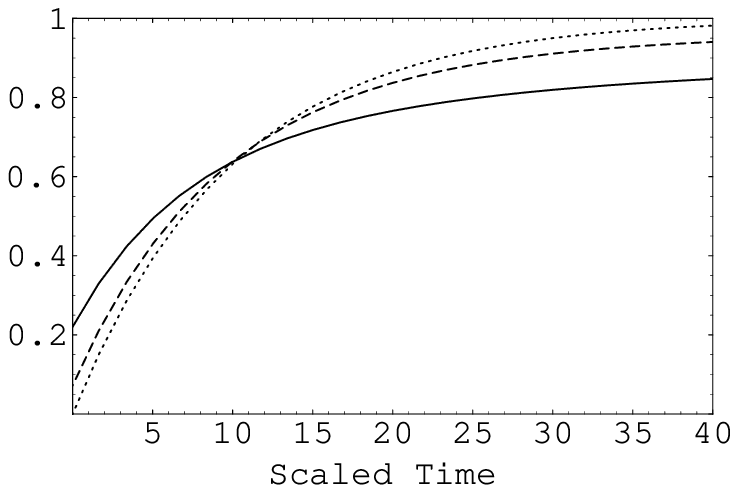}
 \put(-210,80){$\mathcal{D}$}
 \put(-425,80){$\mathcal{D}$}
   \caption{The dot, dash-dot and the solid curves represent the
   Disturbance $\mathcal{D},$
    (a) For the thermal reservoir with
   $\mathcal{N}_1=\mathcal{N}_2=0.00001,0.2,0.6$ (b) For the squeezed
   reservoir with  $\mathcal{M}_1=\mathcal{M}_2=0.001,0.2,0.4$}
  \end{center}
\end{figure}

One says that a system is {\it disturbed} when its initial and
final states  do not coincide. Since the information  is coded on
the input states, then it  may be quantified in terms of
fidelities \cite{Maco}. The closeness of the output quantum state
 $\rho_{f}$ to the input one $\rho_{i}$ is expressed by the quantum fidelity
$\mathcal{F}$, where $0\leq \mathcal{F}\leq 1$ \cite{Ric}.
 Now, we can define the disturbance, $\mathcal{D}$ as
 \begin{equation}
\mathcal{D}=1-\mathcal{F}, \quad
\mathcal{F}=Tr\{\rho_{f}\rho_{i}\}.
 \end{equation}
 Fig(3a),  shows, the behavior of the disturbance of information
 in the case of the thermal reservoir. It is clear that the
 disturbance increases as the scaled time, $\Gamma t$ increases.
 For small values of the thermal photon the disturbance increases
 gradually at the expense of the fidelity of the teleported state.
 For scaled time $\Gamma t\geq 4$, the disturbance
 $\mathcal{D}$ reaches its maximum value. This means that the
 input and the output states are completely different with  the
 entangled state  converted to separable state see(Fig.(2a)).
 As one increases the thermal photon number,  the
 disturbance increases  with time and reaches to a constant value
as soon as the entanglement disappears. So, the entangled state is
completely separable and there is no more information to be
disturbed. For the values which causes a
 sudden-death of entanglement as depicted in Fig.(1a), the
 disturbance sudden be constant. This means at this values  there
 as a sudden death of information.

 Fig.(3b),  describes the behavior of the disturbance of
 information in the presence of the local reservoir. In general,
 the same behavior is seen as that depicted for the thermal
 reservoir, but the disturbance $\mathcal{D}$ increases slowly and
 consequently the  time loss of information is large. This  behavior
 is due to the long-lived entanglement as  seen in Fig.(2b). So,
  for small values of the squeezed reservoir parameters, the time of disturbed
 information can be infinite.

To quantify the amount of information exchange between the state
and the environment during the evaluation, we use the entropy
exchange \cite{Yang},
\begin{equation}
S_e=-Tr\{\rho\log\rho\}.
\end{equation}
 If there is no interaction between the system and its
 surroundings, then the entropy exchange is zero and consequently
 there is no information loss from the system. One can look at the
 environment as an Eavesdropper, who wants to gain more information
 from the entangled system, which carries this information. In
 Fig.(4), we investigate the dynamics of this  phenomenon also for
 the thermal and squeezed environment. In both cases,
  the exchange information increases as one increases the
 reservoir parameters, but as soon as the state turns into a
 separable state, the exchange information decreases. Then there
 is no more interaction with the local environment, therefore the
 exchange information becomes a constant.
\begin{figure}
  \begin{center}
  \includegraphics[width=18pc,height=12pc]{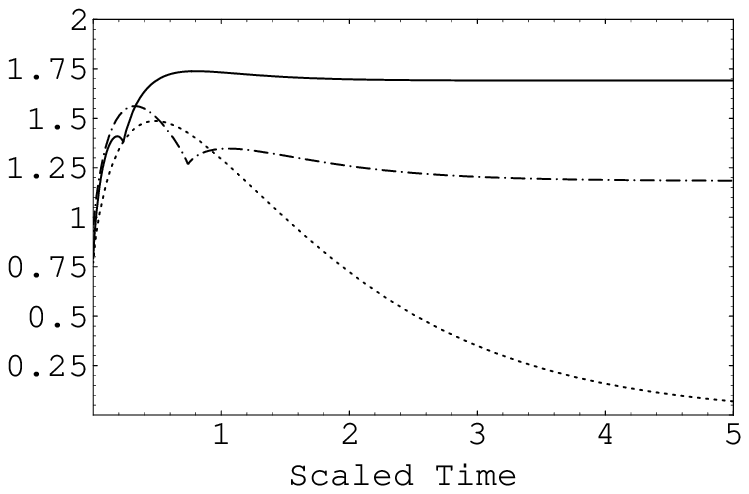}
\includegraphics[width=18pc,height=12pc]{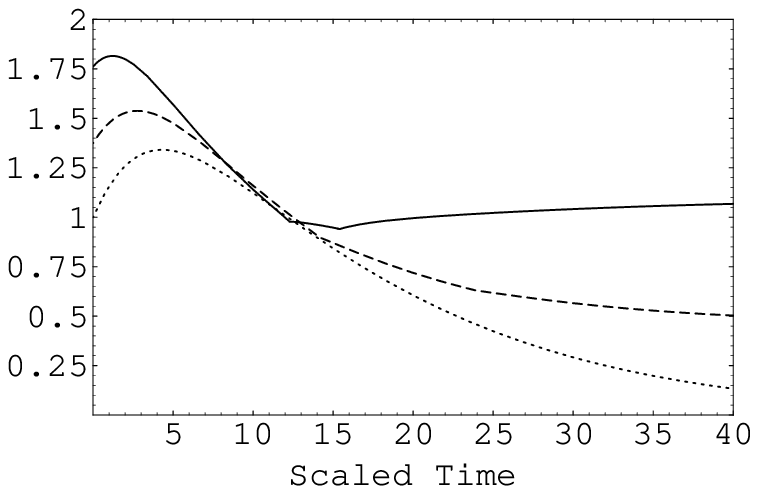}
\put(-210,80){$\mathcal{S}_e$}
 \put(-425,80){$\mathcal{S}_e$}
   \caption{The exchange information, (a) For  the thermal
   reservoir with same values in as Fig.(2a).
    (b) For the squeezed reservoir with same  values as in
    Fig.(2b).}
  \end{center}
\end{figure}
 The effect of the thermal reservoir is seen in Fig.(4a), where
  for small values of the mean photon number
 $\mathcal{N}_1=\mathcal{N}_2=10^{(-4)}$, the exchange entropy increases to reach
 its  maximum values and then decreases with gradually time until it
 reaches to a constant value at $\Gamma t\geq 5$. We  notice
 that, at this time the state turns into  a separable state (see
 Fig.(2a)) so there is no more exchange between the environment
 and the state. As one increases the mean photon number,
 the exchange information becomes constant at $\Gamma
 t\cong 1.3$. For larger value of the mean photon number, say
 $\mathcal{N}_1=\mathcal{N}_2=0.2$, the exchange information
 becomes constant at much earlier time .

The behavior of the exchange information in the  presence of the
local squeezed reservoir is the same as that shown for the thermal
reservoir.  But the time in which the exchange information becomes
constant is much larger than that depicted for the thermal case.
Science  in the squeezed reservoir  case the entanglement is
long-lived.

\section{Conclusion}
In this contribution, we use the Kraus operators to investigate
the dynamics of entangled state passes through a thermal or
squeezed reservoir. The phenomenon  of the entanglement decay and
the sudden death of entanglement are shown for both reservoirs. we
show that  the entanglement lived longer for the squeezed
reservoir. Also, the  disturbance of information is discussed for
both environment, where  it is very sensitive to the thermal
reservoir  parameter much more than the squeezed vacuum reservoir
parameters. For large values of the thermal  photon reservoir, the
information is suddenly disturbed, but it is  disturbed gradually
for the squeezed reservoir. The loss of information is quantified
by the means of the entropy exchange between the environment and
the shared entangled state. For both environments, the exchange
information increases  and leads to a constant  when the system
turns into separable state. For thermal reservoir, the exchange
information becomes constant faster than that for the  squeezed
reservoir.

\end{document}